\documentclass[twocolumn,iop]{openjournal}
\usepackage{amsmath}
\usepackage{amssymb}
\usepackage{multirow}
\usepackage[shortlabels]{enumitem}
\usepackage{hyperref}
\usepackage{longtable}
\usepackage{orcidlink}

\begin{document}

\title{Barycenter kinematics in Local Group analogues}
\author{I. A. López-Paredes\,\orcidlink{0009-0003-0220-8438}}
\affiliation{Escuela de Ingeniería y Ciencias, Tecnológico de Monterrey, Ave. Eugenio Garza Sada 2501, Monterrey 64849, NL, Mexico}
\affiliation{Departamento de Física, Universidad de los Andes, Cra. 1 No. 18A-10 Edificio Ip CP 111711 Bogotá, Colombia}
\affiliation{Dipartimento di Fisica, Università di Roma Tor Vergata, Via della Ricerca Scientifica 1, 00133 Roma, Italy}

\author{J. E. Forero-Romero\,\orcidlink{0000-0002-2890-3725}}
\affiliation{Departamento de Física, Universidad de los Andes, Cra. 1 No. 18A-10 Edificio Ip CP 111711 Bogotá, Colombia; je.forero@uniandes.edu.co}
\affiliation{Observatorio Astronómico, Universidad de los Andes, Cra. 1 No. 18A-10 Edificio H CP 111711 Bogotá, Colombia}

\begin{abstract}
We investigate a new way to identify systems similar to our Local Group (LG) of galaxies in cosmological simulations. We use as an additional constraint the velocity of the LG barycenter, which is measured accurately from cosmic microwave background data.
Using different cosmological simulations we compare pairs selected with this constraint to pairs selected with traditional methods.
Our method selects simulated galaxy pairs where M31's radial approach speed to the Milky Way is $2\%$ to $7\%$ lower depending on the cosmology, while we find no robust change in the tangential velocity.
The difference in the radial component is statistically significant in 86 of the 93 cosmological models we test, but the shift is small (a few km\,s$^{-1}$) compared to the typical velocity dispersion of LG analogues (tens of km\,s$^{-1}$). Matching the observed barycenter velocity therefore has a measurable but modest effect on LG analogue kinematics, and is best treated as an optional refinement, rather than a necessary selection criterion.
\end{abstract}

\begin{keywords}
    {Local Group, N-body simulations, barycenter kinematics}
\end{keywords}

\section{Introduction}
\label{sec:intro}

\begin{table*}[tbp]
    \centering
    \setlength{\tabcolsep}{8pt}
    \caption{Summary of observed reference values. The columns represent, respectively: the physical observable, its measured value, the corresponding units, and the bibliographic source.}
    \begin{ruledtabular}
    \begin{tabular}{ c c c c}  
        Observable & Value & Units & Source\\
        \hline
        $v_b$ & $620\pm15$ & km\,s$^{-1}$ & \cite{planck_collaboration_planck_2020} \\
        $(l,b)_{v_b}$ & ($271.9 \pm 2.0$, $29.6 \pm 1.4$) & deg & \cite{planck_collaboration_planck_2020} \\
        $r_{\text{MW}}$ & $8.29 \pm 0.16 $ & kpc & \cite{mcmillan_mass_2011}\\ 
        $(l,b)_{\text{MW}}$ & $(359.944215 \pm 0.000007,-0.046079 \pm 0.000009)$ & deg &  \cite{yusef-zadeh_position_1999} \\
        $r_{M31}$ & $761\pm11$ & kpc & \cite{li_sub-2_2021} \\
        $(l,b)_{\text{M31}}$ & $(121.174410 \pm 0.000022,-21.572925 \pm 0.000022)$ & deg & \cite{evans_chandra_2010} \\
        $v_{r,\text{M31}}$ & $-109.3\pm4.4$ &  km\,s$^{-1}$ & \cite{van_der_marel_m31_2012} \\
        $v_{t,\text{M31}}$ & $82.4 \pm 31.2$ & km\,s$^{-1}$ & \cite{salomon_proper_2021} \\
        $\mu$ & $-0.88 \pm 0.01$ & -- & This work \\
    \end{tabular}  
    \end{ruledtabular}
    \label{tab:obvs}
\end{table*}

Understanding the orbit between M31 and the Milky Way (MW) is key to explaining how the Local Group (LG) evolved \citep{hartl_milky_2024}. 
Recent research challenges the old view that this orbit is mostly head-on (radial), suggesting instead that it has more sideways motion than previously thought \citep{van_der_marel_first_2019, salomon_proper_2021}. 
These studies rely on cosmological simulations to establish expectations for the tangential velocity components. 
To study LG dynamics, previous works typically identify similar galaxy pairs in simulations by applying specific criteria to halo pairs, including their masses, approach speed, distance, and isolation from other large structures \citep{li_masses_2008,Forero2013, hartl_local_2022, sawala2023, benisty_weighing_2024,benisty_line--sight_2025}.

\begin{figure}[tbp]
    \centering
    \includegraphics[width=0.6\linewidth]{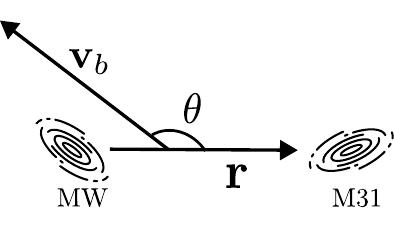}
    \caption{Diagram showing the alignment between key quantities used to select LG analogues. $\mathbf{v}_b$ represents the barycenter velocity of the LG and $\mathbf{r}$ shows the position of M31 relative to the MW.}
    \label{fig:vectors}
\end{figure}

We identified a measurement that previous studies have not used when defining LG-like systems.
This measurement is the LG's barycenter velocity ($\mathbf{v}_b$), which we can determine precisely from cosmic microwave background (CMB) data \citep{planck_collaboration_planck_2020}.
We use this constraint in two ways. First, we consider its magnitude ($v_b$). Second, we examine the direction, measured as the angle ($\theta$) between the barycenter velocity vector and the position vector of M31 relative to the MW ($\mathbf{r}$), as shown in Figure \ref{fig:vectors}.
We express this angle more conveniently as $\mu=\cos\theta$, which comes from the dot product of the unit vectors for $\mathbf{v}_b$ and $\mathbf{r}$.

Previous studies could not use the barycenter velocity because they worked with simulations that were too small in volume. 
Recently, \citet{Forero2022} showed that peculiar velocities can be inaccurate in small simulation boxes due to power spectrum limitations.
Using simulations with volumes ranging from 100 to 2967 Mpc across, they demonstrated that accurate kinematic properties require boxes at least 1 Gpc across.

In this paper, we use simulations to measure how including the barycenter velocity constraint affects the predicted radial and tangential velocities of M31 relative to the MW. We also explore how these effects vary across simulations with different cosmological parameters.

Our paper is organized as follows: Section \ref{sec:obs} covers the observational constraints, and Section \ref{sec:sim} describes the simulations used and the different cosmological models. In Section \ref{sec:LG_anag}, we explain how we select LG analogues using both \textit{traditional} criteria and our new approach. Section \ref{sec:results} presents and discusses our findings, and Section \ref{sec:conclusion} summarizes our conclusions.

\section{Observational constraints}
\label{sec:obs} 

Table \ref{tab:obvs} summarizes the key observational values used to calculate $\mu$. 
We obtain the barycenter velocity of the LG from cosmic microwave background (CMB) measurements \citep{planck_collaboration_planck_2020}. 
The positions of both M31 and the MW are well known from previous studies. 
For the Sun-to-Galactic Center distance, we use the value from \citet{mcmillan_mass_2011}, who derived this by fitting MW mass models to observational data. 
The Galactic Center position comes from VLA measurements reported by \citet{yusef-zadeh_position_1999}. 
For M31, we use the distance measured by \citet{li_sub-2_2021} using Cepheid variable stars, and coordinates from the Chandra Source Catalog \citep{evans_chandra_2010}.

To calculate the position vector of M31 relative to the MW ($\mathbf{r}$) and determine $\mu$, we use a Monte Carlo approach. 
We build a covariance matrix using coordinate measurement error ellipses (from the NASA/IPAC Extragalactic Database) and sample from a multivariate normal distribution. 
The uncertainties in Table \ref{tab:obvs} represent the standard deviations of these samples. 
Our method yields an M31--MW separation of $765 \pm 11$ kpc, matching observations reported by \citet{chamberlain_implications_2022}. 
To compute $\mu$, we generate normal distribution samples for $\mathbf{v}_b$ and the distances to the MW and M31. 
We then compute the dot product between $\mathbf{v}_b$ and $\mathbf{r}$, finding $\mu=-0.88\pm0.01$. 
This result shows a strong anti-alignment between these vectors.

\section{Simulation} \label{sec:sim}
For this study, we use the AbacusSummit simulation suite \citep{AbacusSim, AbacusCode}. 
We analyze the base simulation box with $6912^3$ particles in a 2 $\rm Gpc\,h^{-1}$ volume. 
Because the simulation snapshot is at redshift $z=0.1$, we use the halos' peculiar velocities to project their positions to present-day values. 

We begin with the primary cosmology (c000), which uses the Planck 2018 $\Lambda$CDM model with massive neutrinos. We then analyze 92 additional cosmologies from the sets defined by AbacusSummit, which we group into four families.
The first family (primary and secondary) contains c000, three variants of it with different neutrino treatments (c009 with massless neutrinos, and c019 and c020 with two and zero 60 meV neutrino species), and the four secondary cosmologies (c001--c004) that test different scenarios such as low $\omega_c$, wCDM (c002), high $N_{\text{eff}}$, and low $\sigma_8$.

The second family contains eight of the ten reference cosmologies (c010, c012--c018; c021 and c022 are not included), which match models from other major projects, mostly using massless neutrinos.
For wider parameter testing, the third family (derivative grids) contains the linear derivative grid (c100--c116), with small positive and negative steps in eight parameters, and ten further simulations with smaller spacing (c117--c126). The fourth family contains 50 simulations from the unstructured emulator grid (c130--c180, excluding c159), covering broader parameter ranges.
All cosmologies use the same $\tau$ value (0.0544) and most include 60 meV neutrinos, consistent with Planck 2018 results.

\begin{figure*}[tp]
    \centering
    \includegraphics[width=0.49\linewidth]{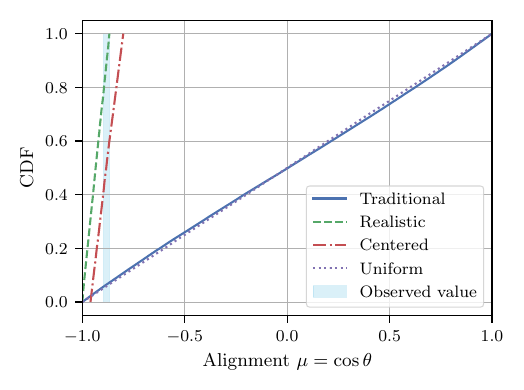}
    \includegraphics[width=0.49\linewidth]{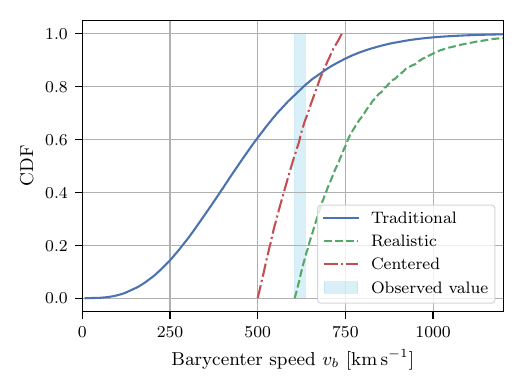}
    \caption{Left panel: cumulative distribution of $\mu$, the cosine of the angle between the barycenter velocity of the LG and the position vector of M31 relative to the MW in the base cosmology (c000). We also plot a uniform distribution from $-1$ to 1 for comparison. Right panel: cumulative distribution of the barycenter speed $v_b$ of the Local Group.
    }
    \label{fig:pmu}
\end{figure*}

\section{Selection of LG analogues}\label{sec:LG_anag}

We use halos identified with the CompaSO Halo Finder \citep{CompaSO}, with selection criteria similar to those in \citet{Forero2022}. 
First, we select halos with maximum circular velocities $V_{\text{max}} > 200$ km\,s$^{-1}$. From these, we identify isolated pairs as two halos, A and B, that are each other's nearest neighbors. 
We also require that no third halo with a maximum circular velocity greater than halo B's can be found within three times the pair's separation distance. 
We define halo A as the one with the greater maximum circular velocity, which we designate as M31. 
This forms our main sample, typically containing about $9\times 10^5$ pairs per simulation box.

From this main sample, we create two subsamples, plus a variant of the second described below. The first subsample, which we call \textit{traditional}, includes pairs meeting these conditions:
\begin{itemize}
    \setlength\itemsep{0em}
    \item Each halo mass between $[0.5, 5.0] \times 10^{12}$ M$_{\odot}$
    \item Separation between halos less than 1 $\rm Mpc\,h^{-1}$
    \item Negative radial relative velocity of M31 with respect to the MW ($v_r < 0$ km\,s$^{-1}$), after accounting for cosmic expansion
\end{itemize}

The second subsample, which we call \textit{realistic}, includes all the \textit{traditional} criteria plus these additional barycenter velocity conditions:
\begin{itemize}
    \setlength\itemsep{0em}
    \item Barycenter speed greater than the observed value, including the uncertainty ($v_b > 605$ km\,s$^{-1}$)
    \item Alignment value less than the observed value, including the uncertainty ($\mu < -0.87$)
\end{itemize}

We use inequalities rather than narrow ranges around the observed values because this approach: (1) captures the observed trends that the barycenter speed is unusually high and the alignment shows strong anti-alignment, and (2) allows us to keep more pairs in our sample that satisfy these conditions. Nonetheless, we introduce for completeness a variation of the \textit{realistic} subsample, hereafter referred to as \textit{centered}, which replaces the one-sided bounds with a symmetric interval around the observed mean values:

\begin{itemize}
    \setlength\itemsep{0em}
    \item 500 km\,s$^{-1}< v_b < 740$ km\,s$^{-1}$
    \item $-0.96<\mu<-0.80$
\end{itemize}

The \textit{traditional} subsample typically contains about $2\times 10^5$ pairs per simulation box, while the \textit{realistic} and \textit{centered} samples contain about $4\times 10^3$ and $6\times 10^3$ pairs, respectively.

\subsection{Statistical tests}

Using the simulations, we compute the cumulative probability distributions for $\mu$ and three key Local Group properties: 
\begin{itemize}
    \setlength\itemsep{0em}
    \item Total Local Group mass ($M_{\text{LG}}$), defined as the sum of the masses of the two halos
    \item Radial velocity of M31 relative to the MW ($v_r$)
    \item Tangential velocity of M31 relative to the MW ($v_t$)
\end{itemize}

To measure how our barycenter velocity constraints affect these properties, we compare the mean values of the \textit{realistic} and \textit{centered} pair distributions with those of the \textit{traditional} pairs.
Because $v_r<0$ for all pairs, we quote the changes in the radial velocity in terms of the radial approach speed $|v_r|$, while the distributions and mean values shown in the figures use the signed $v_r$. This way, a negative difference means that M31 approaches the MW more slowly (less negative $v_r$).
We also use two-sample Kolmogorov--Smirnov (KS) tests to determine if these differences are statistically significant, with $p$-values below 0.05 considered significant. For the base cosmology, we also report the KS statistic $D$, the maximum distance between the two cumulative distributions. Given the large sample sizes, the KS test can detect deviations that are statistically significant but numerically small. We therefore report effect sizes (percentage differences in means, or excess fractions relative to a reference distribution) alongside every significance test, and base our physical interpretation on them rather than on statistical significance alone.

\begin{figure}[htbp]
    \centering
    \includegraphics[width=\linewidth]{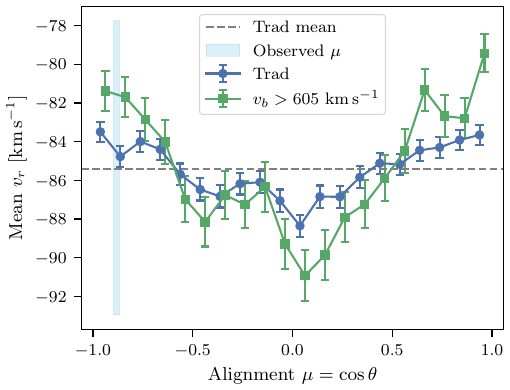}
    \caption{Mean radial velocity of M31 relative to the MW as a function of $\mu=\cos\theta$ in the base cosmology (c000). Blue circles show all \textit{traditional} pairs and green squares show the subset with $v_b>605$ km\,s$^{-1}$. Error bars are the standard error of the mean, and bins with fewer than 20 pairs are omitted. The dashed line marks the mean of the full \textit{traditional} sample.} 
    \label{fig:vr_mu}
\end{figure}

\section{Results} \label{sec:results}

\subsection{Base cosmology}

\begin{figure*}[tbp]
    \centering
    \includegraphics[width=0.9\linewidth]{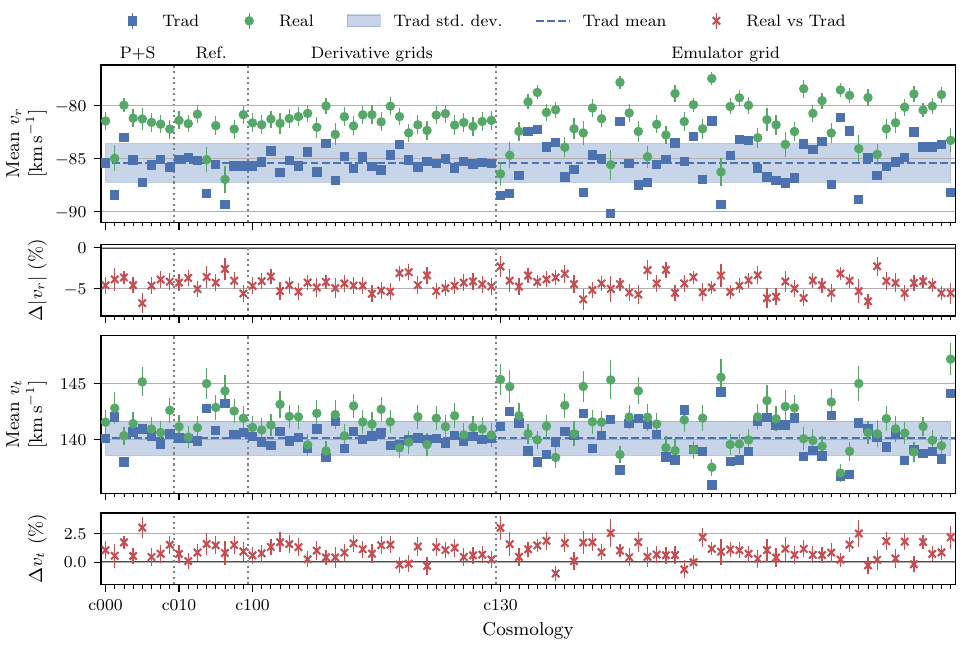}
    \caption{Mean values of M31 velocity distributions for \textit{traditional} and \textit{realistic} pairs in all cosmologies. The first and third panels show the mean $v_r$ and $v_t$; error bars give the standard error of each mean and are smaller than the markers for \textit{traditional} pairs. Dashed lines and shaded regions represent the average and standard deviation of the \textit{traditional} means across all cosmologies. The thin panel below each velocity panel shows the percentage difference between the \textit{realistic} and \textit{traditional} means, with error bars propagated from the standard errors of both means, and a solid line marking zero difference; for $v_r$ this is the difference in $|v_r|$, so negative values indicate a slower radial approach. Cosmologies are ordered by family (primary and secondary, P+S; reference, Ref.; derivative grids; and emulator grid; Section \ref{sec:sim}), separated by dotted lines, and the first model of each family is labeled.}
    \label{fig:mean_diff}
\end{figure*}

Figure \ref{fig:pmu} shows the cumulative distributions of $\mu$ and $v_b$ in the base cosmology (c000), and Figure \ref{fig:vr_mu} shows how the mean $v_r$ depends on $\mu$ in the same cosmology.
The $\mu$ distribution of \textit{traditional} LG pairs deviates only mildly from uniform, with a modest excess of pairs at high $|\mu|$: $11.1\%$ of pairs have $|\mu| > 0.9$,  compared to the $10.0\%$ expected under a uniform distribution.
This deviation is formally significant ($N = 211{,}263$; KS test, $D = 0.013$, $p\approx 10^{-29}$),  but the magnitude of the effect is small. 
We find no meaningful asymmetry between aligned and anti-aligned pairs ($50.2\%$ with $\mu > 0$ versus $49.8\%$ with $\mu < 0$; mean $\mu = 0.002$).

We find statistically significant differences between \textit{traditional} and \textit{realistic} pairs in M31's radial velocity distribution ($N=211{,}263$ \textit{traditional} and $N=3983$ \textit{realistic} pairs; $D=0.035$ and $p\approx10^{-4}$). 
The mean radial approach speed $|v_r|$ changes by $-4.62\%$, so M31 approaches the MW more slowly ($|v_r|=81.5\pm0.9$ km\,s$^{-1}$ for \textit{realistic} pairs, compared with $85.4\pm0.1$ km\,s$^{-1}$ for \textit{traditional} pairs). The differences in M31's tangential velocity distribution and total mass ($M_{\text{LG}}$) are not statistically significant.

To identify which selection cut drives the shift in $v_r$, we examine the mean $v_r$ of the \textit{traditional} sample as a function of $\mu$ (Figure \ref{fig:vr_mu}). 
The profile is U-shaped and approximately symmetric about $\mu=0$; pairs that are strongly aligned or anti-aligned with $\mathbf{v}_b$ have a less negative $v_r$ than pairs with $\mu\approx0$.
Restricting the sample to $v_b>605$ km\,s$^{-1}$ increases the amplitude of this dependence.

Applying each barycenter constraint separately, $v_b>605$ km\,s$^{-1}$ alone changes the mean $|v_r|$ by only $-0.46\%$ ($-0.4$ km\,s$^{-1}$), while $\mu<-0.87$ alone gives $-2.1\%$ ($-1.8$ km\,s$^{-1}$). 
Applying both gives the aforementioned shift of $-4.62\%$ in the \textit{realistic} pairs ($-3.9$ km\,s$^{-1}$), which is larger than the sum of the individual shifts. 
This interaction is also visible within the strongly anti-aligned subsample. We perform a bootstrap analysis with 5000 resamples and find that for $\mu<-0.87$, pairs with $v_b>605$ km\,s$^{-1}$ have a mean $|v_r|$ that is lower by $3.0\pm1.0$ km\,s$^{-1}$ than those with $v_b\leq605$ km\,s$^{-1}$ ($\approx3\sigma$).
The shift between \textit{realistic} and \textit{traditional} pairs therefore reflects a dependence of $v_r$ on the alignment between $\mathbf{v}_b$ and $\mathbf{r}$, enhanced at high $v_b$. The effect remains small compared with the dispersion of $v_r$ ($\approx57$ km\,s$^{-1}$). 

Note that because the profile is approximately symmetric in $\mu$, the dependence of $v_r$ on the alignment is not specific to anti-alignment. 
The \textit{realistic} selection ($\mu<-0.87$) isolates the anti-aligned branch because this is where the observed Local Group lies ($\mu=-0.88\pm0.01$, Section \ref{sec:obs}).

\subsection{All cosmologies}

\begin{table*}[tp]
\centering
\caption{Percentage differences in the distributions of the M31 velocity components ($v_r$ and $v_t$) and the total mass of the Local Group ($M_{\text{LG}}$). 
For each observable and sample type, the table reports the mean difference relative to the traditional sample, computed (I) using only the cosmologies where the difference is statistically significant and (II) using all 93 cosmologies. For the radial velocity, the differences refer to the approach speed $|v_r|$. The last column gives the number of cosmologies where the KS test indicates a significant difference between the distributions.}

\begin{ruledtabular}
\begin{tabular}{llccc}
Observable 
& Sample 
& Mean difference I (\%) 
& Mean difference II (\%)  
& Significant cosmologies (out of 93) \\
\hline
$|v_r|$
& Realistic
& $-4.64$
& $-4.51$ 
& 86 \\
& Centered
& $-2.84$
& $-2.14$ 
& 37 \\
\hline
$v_t$ 
& Realistic 
& $1.63$ 
& $0.96$ 
& 15 \\
& Centered
& $-0.75$ 
& $-0.34$
& 12 \\
\hline
$M_{\text{LG}}$ 
& Realistic 
& $-0.83$
& $-0.32$
& 9 \\
& Centered
& --
& $-0.07$
& 0
\end{tabular}
\end{ruledtabular}
\label{tab:results}
\end{table*}

\begin{figure}[tbp]
     \centering
     \includegraphics[width=\linewidth]{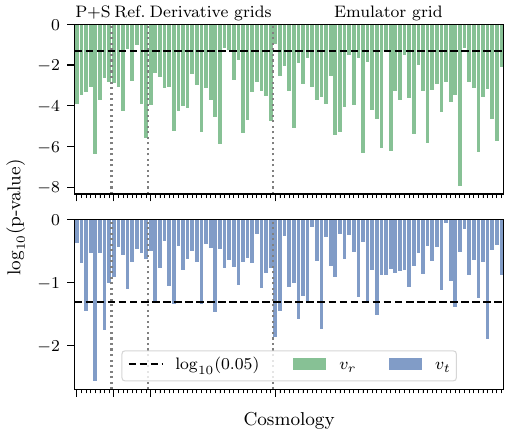}
     \caption{Statistical significance of the differences between \textit{realistic} and \textit{traditional} selection criteria for radial velocity $v_r$ (top panel) and tangential velocity $v_t$ (bottom panel). The $p$-values shown are from two-sample Kolmogorov--Smirnov tests across all cosmologies, grouped by family as in Figure \ref{fig:mean_diff}.
     }
     \label{fig:pvalues}
\end{figure}

\begin{figure*}[tp]
    \centering
    \includegraphics[width=0.9\linewidth]{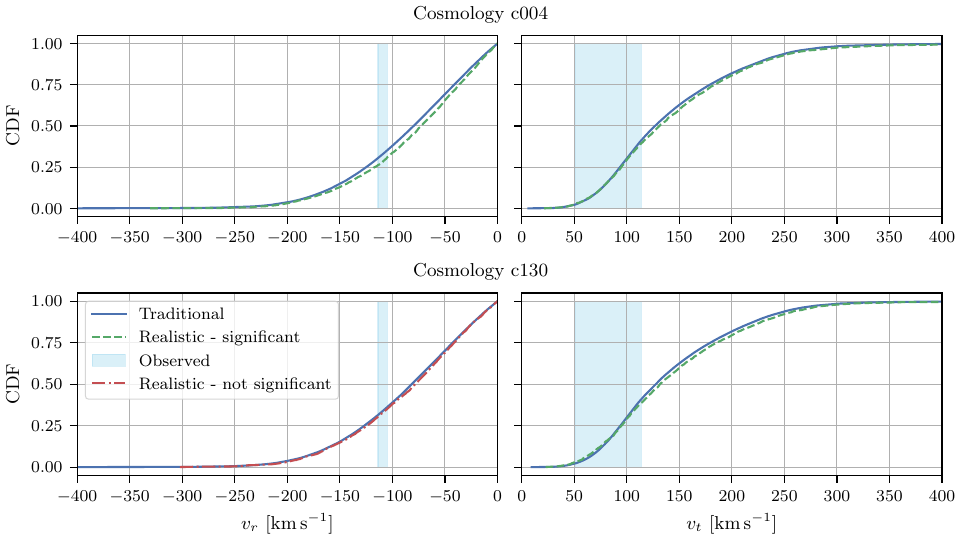}
    \caption{Cumulative distributions of M31 velocities relative to the MW in cosmologies c004 and c130. Cosmology c004 has the largest difference in the mean between \textit{traditional} and \textit{realistic} pairs in both components, and c130 has a nearly identical shift in $v_t$. In cosmology c004, the difference in both $|v_r|$ ($-7\%$) and $v_t$ (3\%) was statistically significant. In cosmology c130, only the difference in $v_t$ (3\%) was statistically significant.}
    \label{fig:cdf_compare}
\end{figure*}

Our barycenter velocity constraints systematically shift the radial velocity distribution of M31 in nearly all of the cosmological models we test, while the effect on the tangential velocity is weaker and not consistent in sign. Figure \ref{fig:mean_diff} shows how the mean values of tangential and radial velocities change across all cosmologies. 
\textit{Realistic} pairs show a lower radial approach speed $|v_r|$ (less negative $v_r$) in all cosmologies, with mean differences between $-2\%$ and $-7\%$, although the difference is significant in 86 of the 93 cosmologies according to the KS tests (Figure \ref{fig:pvalues}).
On the other hand, $v_t$ tends to increase for \textit{realistic} pairs. 
The effect in this case is weaker, with changes between $-1\%$ and $3\%$, and significant differences in only 15 of the 93 cosmologies. 
The effect on total mass ($M_{\text{LG}}$) is significant in only 9 cosmologies. 

We compute the average change in the observables across cosmologies with significant differences, and, to avoid selection bias, we repeat the procedure across all cosmologies. The results, presented in Table \ref{tab:results}, further show the trend described above. The change in $|v_r|$ seen in the \textit{realistic} sample is the largest and most robust across cosmologies. The mean change of roughly $-4.6\%$ in the significant subset remains almost the same ($-4.5\%$) when averaged over all cosmologies. The \textit{centered} sample follows the same trend but with smaller magnitudes. Tangential velocity behaves differently. The mean difference in significant cosmologies is small (around 1.6\% for the realistic sample), and when averaged over all cosmologies the value drops below 1\%. For the \textit{centered} sample, the average change is negative (which corresponds to smaller velocities), but the magnitude is below 1\% for both averages (across significant cosmologies and across all cosmologies). $M_{\text{LG}}$ has the weakest effect, with changes below 1\% and significant differences in only 9 cosmologies for the \textit{realistic} sample and none for the \textit{centered}.

The cosmology with the biggest differences in both $|v_r|$ and $v_t$ between \textit{traditional} and \textit{realistic} pairs is c004 (a baseline $\Lambda$CDM with low $\sigma_8=0.75$). In this cosmology, the difference is statistically significant for both observables, with mean differences of $-7\%$ in $|v_r|$ and 3\% in $v_t$. Cosmology c130 (an emulator-grid cosmology around the baseline) shows a nearly identical shift in $v_t$ (3\%), but in this case the difference is significant only in $v_t$. The corresponding cumulative distributions are shown in Figure \ref{fig:cdf_compare}.

As c004 and c130 illustrate, a given cosmology can show significant differences in some observables but not in others. Table \ref{tab:combinations} summarizes, for each cosmology, which combinations of observables display statistically significant differences between \textit{traditional} and \textit{realistic} or \textit{centered} pairs. For the \textit{realistic} sample, most cosmologies (65) are affected in $v_r$ alone, and only 2 show significant differences in all three observables; for the \textit{centered} sample, half of the cosmologies (47) show no significant difference at all.

\begin{table}[htbp]
    \centering
    \caption{Number of cosmologies (out of 93) where the difference between \textit{traditional} pairs and \textit{realistic} or \textit{centered} pairs is statistically significant, for each combination of observables.}
    \begin{ruledtabular}
    \begin{tabular}{ccc}
    Observables & Realistic & Centered \\
    \hline
    None  &  6 & 47\\
    $v_r$ only & 65 & 34 \\
    $v_t$ only & 1 & 9 \\
    $v_r$ and $v_t$ & 12 & 3 \\
    $M_{\text{LG}}$ only & 0 & 0\\
    $v_r$ and $M_{\text{LG}}$ & 7 & 0 \\
    $v_t$ and $M_{\text{LG}}$ & 0 & 0\\
    All three & 2 & 0\\
    \hline
    Total & 93 & 93
    \end{tabular}
    \end{ruledtabular}
    \label{tab:combinations}
\end{table}

As for the base cosmology, although the barycenter-based selection produces mean shifts that are statistically significant given the large sample sizes, the absolute shifts ($\sim$ a few km\,s$^{-1}$) are small compared to the typical velocity dispersion of LG analogues (tens of km\,s$^{-1}$). 

\begin{figure*}[tp]
    \centering
    \includegraphics[width=0.49\linewidth]{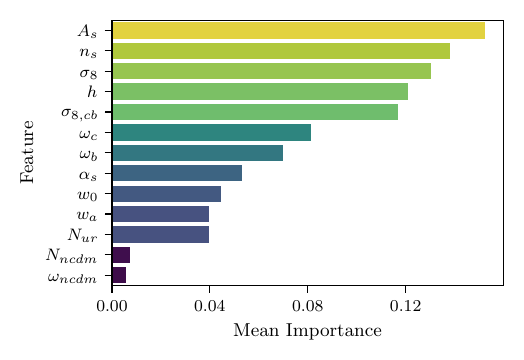}
    \includegraphics[width=0.49\linewidth]{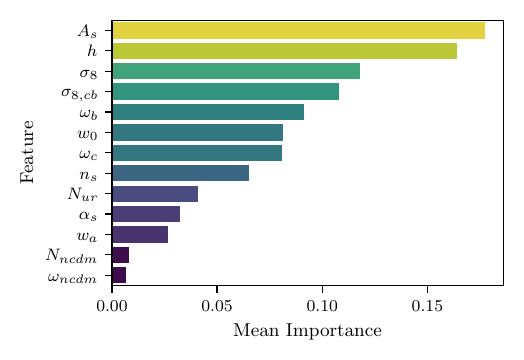}
    \caption{Relative importance of various cosmological parameters for predicting whether the difference in M31's velocities between \textit{centered} and \textit{traditional} pairs is statistically significant as measured by the KS test ($p<0.05$). The left panel shows the parameter importance for $v_r$, and the right panel for $v_t$. Importance is calculated using a Random Forest algorithm.}
    \label{fig:feature}
\end{figure*}

\subsection{Cosmological parameters}

\begin{figure*}[tp]
    \centering
    \includegraphics[width=0.49\linewidth]{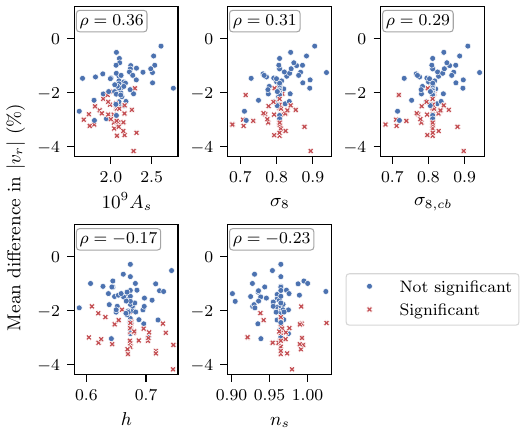}
    \includegraphics[width=0.49\linewidth]{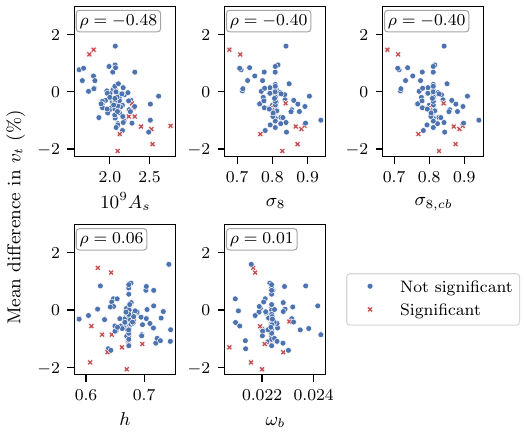}
    \caption{Scatter plots of the mean difference in M31's velocities between \textit{centered} and \textit{traditional} pairs versus the most important cosmological parameters for predicting whether a cosmology leads to significant differences. The left panel corresponds to $|v_r|$, and the right panel to $v_t$. Each point is one cosmology; red crosses mark cosmologies where the KS test finds a statistically significant difference ($p<0.05$) between \textit{centered} and \textit{traditional} pairs in that velocity component, and blue circles those where it does not. For each parameter, $\rho$ is the calculated Spearman correlation coefficient with respect to the mean difference in the velocity.}
    \label{fig:scatter}
\end{figure*}

To explore which cosmological parameters determine whether the velocity distributions of \textit{traditional} and \textit{realistic} pairs differ significantly, we trained a Random Forest classifier on the 93 AbacusSummit cosmologies, using the KS-test outcome (significant or not, at $p<0.05$) as the target. Under the \textit{realistic} filter, however, the two classes are strongly imbalanced: the difference is significant in 86 of 93 cosmologies for $v_r$ and in only 15 for $v_t$. For $v_r$, the classifier reaches a macro F1 score of 0.48 and a weighted F1 score of 0.89. The \textit{realistic} sample therefore contains too few minority-class cosmologies for this analysis.

We therefore repeated the analysis with the \textit{centered} filter, which selects pairs in a window around the mean observed values of $v_b$ and $\mu$. For $v_r$ this gives a more balanced split (37 significant and 56 non-significant cosmologies), and the classifier reaches a macro F1 score of 0.55 and a weighted F1 score of 0.58. For $v_t$ the classes remain imbalanced (12 significant and 81 non-significant cosmologies); the classifier reaches a macro F1 score of 0.70 and a weighted F1 score of 0.87. The resulting importances (Figure \ref{fig:feature}) should thus be read as indicative rather than definitive, particularly for $v_t$. For both velocity components, the most important parameter is the amplitude of scalar perturbations, $A_s$. Because the target is the outcome of a KS test, these importances indicate which parameters influence the \emph{statistical detectability} of a difference at the sample size available in each cosmology, not the \emph{magnitude} of the shift.

To address the magnitude directly, we computed Spearman correlation coefficients between the mean percentage difference in each velocity component and the individual cosmological parameters (Figure \ref{fig:scatter} and Table \ref{tab:spearman}). Under the \textit{centered} filter, the difference in $|v_r|$ correlates positively with $A_s$ and $\sigma_8$, while the difference in $v_t$ correlates negatively with both.
Under the \textit{realistic} filter, the $v_t$ correlations are similar, but $|v_r|$ shows no measurable correlation with either parameter. This may reflect the nearly universal significant difference in $|v_r|$ under the \textit{realistic} filter.
Since the differences in $|v_r|$ are negative, the positive correlation means that the reduction in radial approach speed is smaller in cosmologies with higher $A_s$ (and $\sigma_8$). For $v_t$, the negative correlation means the change in tangential velocity becomes more negative at higher $A_s$.
Note that $A_s$ and $\sigma_8$ are not independent; in most AbacusSummit cosmologies $\sigma_8$ is the design parameter and $A_s$ is rescaled to match it ($\sigma_8^2 \propto A_s$ at fixed other parameters). The two correlations should therefore not be read as separate results.

\begin{table}[htbp]
    \centering
    \caption{Spearman correlation coefficients $\rho$, with their $p$-values in parentheses, between the mean percentage difference in each velocity component (relative to the \textit{traditional} sample) and the cosmological parameters $A_s$ and $\sigma_8$, across the 93 cosmologies.}
    \setlength{\tabcolsep}{3pt}
    \begin{ruledtabular}
    \begin{tabular}{llcc}
    Sample & Component & $A_s$ & $\sigma_8$ \\
    \hline
    Centered  & $|v_r|$ & $0.36$ ($4.0\times10^{-4}$)  & $0.31$ ($2.2\times10^{-3}$) \\
              & $v_t$ & $-0.48$ ($8.6\times10^{-7}$) & $-0.40$ ($5.8\times10^{-5}$) \\
    Realistic & $|v_r|$ & $0.04$ ($0.74$)              & $0.11$ ($0.28$) \\
              & $v_t$ & $-0.35$ ($6.1\times10^{-4}$) & $-0.45$ ($5.7\times10^{-6}$) \\
    \end{tabular}
    \end{ruledtabular}
    \label{tab:spearman}
\end{table}

Taken together, the prominence of $A_s$ in the classification and the consistent correlations with $A_s$ and $\sigma_8$ suggest that the amplitude of matter fluctuations may influence how strongly barycenter kinematics affect the velocities of LG analogues. However, testing which physical mechanism is responsible for this effect requires a dedicated physical study beyond the scope of this work. These results should therefore be regarded as exploratory, identifying statistical trends rather than establishing physical causation.

\section{Conclusion}\label{sec:conclusion}
In this work, we explored the barycenter velocity ($\mathbf{v}_b$) as a new kinematic constraint for defining Local Group (LG) analogues.
This quantity has very low observational uncertainty but was not included in previous studies due to the need for sufficiently large simulation volumes to achieve numerical convergence in peculiar velocities \citep{Forero2022}.
Using simulations from the AbacusSummit project, with a box size of 2 $\rm Gpc\,h^{-1}$, we have shown that including this constraint introduces a small but systematic shift in the predicted radial velocity distribution of M31.

We demonstrated this effect by comparing LG analogues selected using our proposed constraints (\textit{realistic} pairs: $v_b > 605$ km\,s$^{-1}$ and $\mu < -0.87$) with those selected using \textit{traditional} constraints (isolated pairs with each halo mass between $0.5$ and $5.0\times10^{12}$ M$_\odot$, separation distance less than 1 Mpc $h^{-1}$ and negative radial velocity).
We also considered a \textit{centered} variant with symmetric windows around the observed values of $v_b$ and $\mu$.
Our results show that in the \textit{realistic} pairs, M31 tends to have a smaller radial approach speed $|v_r|$ (less negative $v_r$) compared to the \textit{traditional} pairs, while we find no robust change in its tangential velocity $v_t$.
We quantified these effects using the difference in means and assessed their significance through two-sample Kolmogorov--Smirnov tests.

The radial effect is present in nearly all cosmological models. In our analysis of 93 different cosmologies, we found significant differences in the radial velocity distribution for 86 models, while the tangential velocity distribution showed significant differences in only 15 cosmologies.
The mean difference between \textit{realistic} and \textit{traditional} pairs was $-4.5\%$ in $|v_r|$ (a slower radial approach), averaged across all cosmologies. However, this shift amounts to only a few km\,s$^{-1}$ (a reduction of 3.9 km\,s$^{-1}$ in $|v_r|$ in the base cosmology), which is small compared to the typical velocity dispersion of LG analogues ($\approx57$ km\,s$^{-1}$ in the base cosmology).
For $v_t$, the mean difference averaged across all cosmologies is $+1\%$ under the \textit{realistic} filter but changes sign ($-0.3\%$) under the \textit{centered} filter; we therefore do not consider the tangential velocity result a robust constraint.
The total mass $M_{\text{LG}}$ (defined in this work as the sum of the masses of the Milky Way and M31) is essentially unaffected, with changes below 1\%.

We also found indications that $A_s$ is the cosmological parameter most predictive of the statistical detectability of differences in the velocity distributions, and that the size of these differences correlates with $A_s$ and $\sigma_8$.
However, these results are exploratory, and further work is needed to understand the influence of cosmological parameters on LG kinematics. 
Similarly, while we have demonstrated statistical differences in velocity distributions, a deeper exploration of the physical mechanisms explaining why barycenter kinematics correlate with orbital properties remains for future work. 
Assessing whether these results change our understanding of the Local Group properties and history ---for instance in light of studies such as \cite{chamberlain_implications_2022, hartl_milky_2024, benisty_weighing_2024, benisty_line--sight_2025}--- is also left for future work.

In summary, our results show that the kinematic constraints proposed in this work have a statistically significant but small effect on the radial velocity of M31, present in nearly all cosmological models, and no robust effect on its tangential velocity. Given the modest size of this effect relative to the velocity dispersion of LG analogues, there may be no need to take this constraint into account when defining LG analogues. Imposing it reduces the number of available analogues by a factor of $\sim50$ for the \textit{realistic} sample, which can make the scientific interpretation of other properties of the MW--M31 system harder, without a correspondingly strong gain in discriminating power. We therefore suggest treating the barycenter constraint as an optional refinement, whose use should be assessed on a case-by-case basis depending on the science question at hand.

\section*{Acknowledgments}

We thank the anonymous referees for their careful reading of the manuscript and for comments that helped improve this work.

This work made use of the NASA/IPAC Extragalactic Database (NED), funded by the National Aeronautics and Space Administration and operated by the California Institute of Technology. 

\bibliography{refs}
\bibliographystyle{aasjournal}

\end{document}